\documentclass{ecai} 

\usepackage{latexsym}
\usepackage{amssymb}
\usepackage{amsmath}
\usepackage{amsthm}
\usepackage{booktabs}
\usepackage{enumitem}
\usepackage{graphicx}
\usepackage{color}
\usepackage{subcaption}
\usepackage[hidelinks]{hyperref}
\usepackage{xr} 
\newcommand{\BibTeX}{B\kern-.05em{\sc i\kern-.025em b}\kern-.08em\TeX}

\DeclareMathOperator*{\argmax}{arg\,max}

\begin{document}


\begin{frontmatter}


\paperid{1348} 


\title{Emergence of Fair Leaders via Mediators in Multi-Agent Reinforcement Learning}


\author[A]{\fnms{Akshay}~\snm{Dodwadmath}\thanks{Corresponding Author. Email: akshay.dodwadmath@ruhr-uni-bochum.de.}}
\author[A]{\fnms{Setareh}~\snm{Maghsudi}}

\address[A]{Ruhr University Bochum, Germany}


\begin{abstract}
Stackelberg games and their resulting equilibria have received increasing attention
in the multi-agent reinforcement learning literature. Each stage of a traditional Stackelberg game involves a leader(s) acting first, followed by the followers. In situations where the roles of leader(s) and followers can be interchanged, the designated role can have considerable advantages, for example, in first-mover advantage settings.
Then the question arises:  \textit{Who should be the leader and when}? A bias in the leader selection process can lead to unfair outcomes. This problem is aggravated if the agents are self-interested and care only about their goals and rewards. We formally define this leader selection problem and show its relation to fairness in agents' returns. Furthermore, we propose a multi-agent reinforcement learning framework that maximizes fairness by integrating mediators. Mediators have previously been used in the simultaneous action setting with varying levels of control, such as directly performing agents' actions or just recommending them. Our framework integrates mediators in the Stackelberg setting with minimal control (leader selection). We show that the presence of mediators leads to self-interested agents taking fair actions, resulting in higher overall fairness in agents' returns.
\end{abstract}
\end{frontmatter}

\section{Introduction}
\label{section:Introduction}
Stackelberg games involve hierarchical decision-making where the leader acts first, followed by the followers, who act by observing the realization of the leader's strategy. There can be considerable advantages to being a leader indefinitely in these games, especially in games with first-mover or leadership advantages. Examples include quantity competition
with substitute goods, where the leader earns higher profits than a follower \citep{Endogenizing2}, 
or a multi-agent traffic control system, where the leader gets a higher priority and can dominate the decision-making process and the corresponding traffic \citep{traffic_routes}. In the context of multi-agent reinforcement learning (MARL), 
many future applications are predicted to be of mixed-motive or purely competitive nature \cite{openproblemscooperativeai}, which can naturally result in leaders taking unfair and selfish actions, benefiting their purposes, leading to high disparity in gains. Hence, it is essential to provide incentives for leaders to take fair actions instead, which can bring about improved equity. 

In Stackelberg games with leadership advantage, having a dynamic order of agents acting as leaders can indirectly result in such an incentive. For example, previous works have shown that alternating agents as leaders make them take fair actions instead of selfish ones in non-episodic settings \citep{alternatingPD}\citep{PDwithoutSynchrony}. In this scenario, for any agent acting as a leader, there is always an incentive that other agents will reciprocate with their own fair leader action in subsequent steps. However, the same cannot be said about episodic settings since there is always a chance of defection at the terminal states, which can result in a cascading effect of defections in all states \citep{marl-book}; in other words, there is a loss of incentive for leaders to take fair actions across an episode. Moreover, having rigid rules of leader selection, such as alternation, is not optimized for fairness and can bring about sub-optimal overall fairness.
Instead, we propose introducing mediators that can dynamically select leaders, with the objective of optimizing overall fairness. Further, we show that integrating mediators for leader selection can intrinsically incentivize leaders to take fair actions.  

A closely related set of works that also consider dynamic leader selection are those that allow agents to select leaders on their own; for example, through a pre-agreement stage or voting \citep{BeALeaderOrFollower}\citep{traffic_routes}\citep{who_leads}\citep{vote_based}. However, these works assume a level of cooperation or the agents reaching an agreement on how the leaders are selected, but in a non-cooperative context, this might not always be the case. In contrast, by delegating the leader selection process to a central trusted entity \citep{openproblemscooperativeai} such as a mediator, there is always a consensus. Further, self-interested agents do not generally have an incentive to consider fairness while selecting leaders.

Mediators have been discussed before in simultaneous action games. The simplest form of mediator captures the notion of correlated equilibrium \citep{AUMANN}. Monderer and Tennenholtz \citep{MONDERER2009} introduced
a powerful form of mediator, one that can act on behalf of the agents. Ivanov et al. \citep{mediatedRL} extend this to the RL setting by introducing Markov mediators, which are separate agents with pre-defined goals. Other forms of mediators include those that have an indirect influence on agents, such as through monetary payments or contracts \citep{kimplementation}\citep{principalagentreinforcementlearning}. We introduce Markov mediators in the Stackelberg setting with dynamic leaders. In particular, we define mediators that constrain the leader selection process such that fairness is maximized among RL agents. To the best of our knowledge, ours is the first work that integrates mediators in the Stackelberg context.

Emergent prosocial behavior is essential for successfully integrating AI agents into human society \cite{LoLA}. Our work aims to take a step towards achieving this by showing that the presence of central entities such as mediators leads to the emergence of fair agents, as opposed to selfish ones without it.

\subsection{Our contributions}
We first introduce a new setting of Markov Stackelberg games with dynamic leaders. We formally define this model for decentralized systems where each agent independently learns and makes decisions. Next, we demonstrate how fairness is related to the order in which agents act as leaders in this setup and define fair Markov mediators that can be integrated as central entities to determine this order.  

We then focus our attention on analyzing the problem within the context of multi-agent reinforcement learning. To achieve this, we assume agents learn with standard Q-learning and define corresponding Q-functions adapted to this setting. We also define Q-functions for mediators, specifically designed to optimize overall fairness, and propose a joint agents-mediator framework (JAM-QL). We first investigate the theoretical properties of this framework in the full-information setting, assuming the agents and mediator have complete knowledge about the environment dynamics. Notably, we establish the convergence of the agents and mediator to optimally fair policies under certain conditions. Next, we empirically validate our framework across multiple environments, adapted to our problem setting. We show that our mediator-integrated framework leads to the emergence of fair leaders and demonstrates higher levels of fairness against different baselines. Finally, we adopt a version of the framework to deep RL settings using neural networks, making it scalable to settings with high-dimensional inputs and complex environments. 

\section{Preliminaries}

\subsection{Markov games}

An $N$-player Markov game (or stochastic game)  can be defined by a tuple $\Gamma \triangleq\left\langle\mathcal{I}, \mathcal{S},\left\{\mathcal{A}_i\right\}_{i \in \mathcal{I}}, \mathcal{P},\left\{r_i\right\}_{i \in \mathcal{I}}, \gamma\right\rangle. \; \mathcal{I}=\{1,2, \ldots, N\}$ denotes the set of agents and $s \in \mathcal{S}$ is the global state set of the environment. $\mathcal{A}_i$ denotes the action space of agent $i$ and the joint action space $\mathcal{A}=\prod_{i=1}^N \mathcal{A}_i$ is the product of action spaces of all agents. $\mathcal{P}: \mathcal{S} \times \mathcal{A} \rightarrow \mathrm{\Omega}(\mathcal{S})$ denotes state transition function, where $\mathrm{\Omega}(X)$ is the set of probability distributions in $X$ space. $r_i: \mathcal{S} \times \mathcal{A} \rightarrow {R}$ is the reward function of agent $i$ and $\gamma$ is the discount factor. At time step
$t$, each agent chooses an action $a^t_i \in A_i$ at state $s^t \in S$ based
on its own policy $\pi_i: \mathcal{S} \rightarrow \mathrm{\Omega}(A_i)$ and receives feedback in the form of $r_i\left(s^t, \boldsymbol{a}^{t}\right)$, where $\boldsymbol{a}^{t}=\left(a^t_1, \ldots, a^t_N\right) \in \mathcal{A}$. The environment moves to a new state $s^{t+1} \sim \mathcal{P}\left(s^{t+1} \mid s^t, \boldsymbol{a}^{t}\right)$ as a result of joint action $\boldsymbol{a}^{t}$. If ${\boldsymbol{\pi}_{-i}} $ is the joint policy of all agents other than $i$, the value function of agent $i$ is given by $V_i^{\pi_i}(s)=\mathbb{E}_{\pi_i \mid \boldsymbol{\pi}_{-i}}\left[\sum_t \gamma^{t} r_i\left(s^t, \boldsymbol{a}^t\right) \mid s^0 = s\right]$ and the 
optimal value function is given by $V_i^* = \max_{\pi_i} V_i^{\pi_i}(s)$ for all $s \in \mathcal{S}$. The action-value function is given by
$
Q_i^{\pi_i}(s, a_i)=\mathbb{E}_{\pi_i \mid \boldsymbol{\pi}_{-i}}\left[\sum_t \gamma^{t} r_i\left(s^t, \boldsymbol{a}^t\right) \mid s^0 = s, a^0 = a\right]
$ and the optimal action-value function is given by $Q_i^*(s,a_i) = \max_{\pi_i} Q_i^{\pi_i}(s,a_i)$ for all $s \in \mathcal{S}$ and $a_i \in \mathcal{A}_i$. The optimal policy at any state can be derived from the optimal Q-function using $\pi_i^*(s) = \argmax_{a_i} Q_i^*(s,a_i)$.

\subsection{Markov Stackelberg games with dynamic leaders}
A Stackelberg model dictates a hierarchy in which the agents act. Consider a two-player Markov game with the standard Stackelberg model. The standard model has a fixed leader and a follower at each state $s \in S$. The leader enforces its strategy $\pi_1: \mathcal{S}  \rightarrow \mathrm{\Omega}\left(\mathcal{A}_1\right)$ to the follower, whose observation space will now include the leader's behavior  $\pi_{2}: \mathcal{S} \times \mathcal{A}_1 \rightarrow \mathrm{\Omega}\left(\mathcal{A}_2\right)$, where $\pi_i \in \mathrm{\Pi}_i$ and $\mathrm{\Pi}_i$ represents the policy space. In a Markov Stackelberg game with dynamic leaders, the roles of leader and follower can be interchanged at each state. Then the strategy of each agent includes both the leader and follower observation space, $\pi_{1}: \mathcal{S}  \ \cup  \ \mathcal{S}  \times \mathcal{A}_2 \rightarrow \mathrm{\Omega}\left(\mathcal{A}_1\right)$ and similarly for the other agent. We can similarly extend the model to the $N$-agent case, where each agent's observation space includes the behavior of all possible leaders.

\begin{figure}[t]
    \centering
    \includegraphics[width=1.0\linewidth]{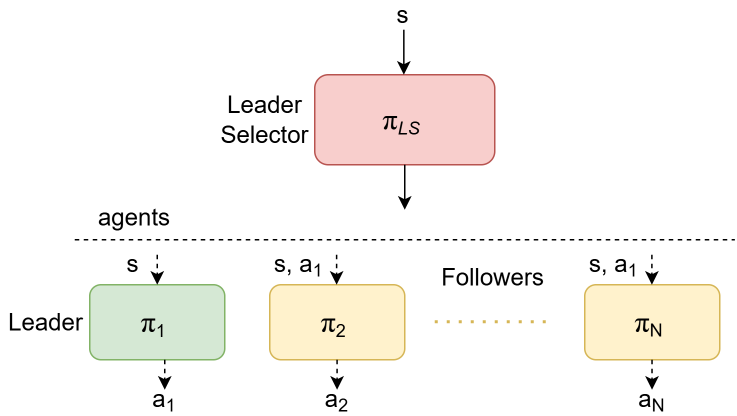}
    \caption{One stage of a Stackelberg game with dynamic leaders. Given a Markov state $s$, a leader selection strategy $\pi_{LS}$ selects a leader, based on which the agents play a single-stage Stackelberg game at that state.}
    \vspace{15pt}
    \label{fig:perstateSG}
\end{figure}

For such games, a leader selection policy $\pi_{LS}$ defines a multi-agent MDP for the agents, and each state of the MDP can be analyzed as a single-stage Stackelberg game between the agents. We provide an overview of this process in Figure \ref{fig:perstateSG}.
Further, for a given $\pi_{LS}$, we can employ the general solution concept of Markov Perfect Equilibrium (MPE) for the agents. At MPE, each agent best responds to the equilibrium strategy of the other agents, and no agent can improve their payoff by unilaterally deviating in any state. For the $N$-agent case, that means
\[
\begin{aligned} 
\forall s \in \mathcal{S}, \forall i \in \mathcal{I}, \forall \pi_i \in \mathrm{\Pi}_i,\\
V^{\pi^{\text{eq}}_i}_i \left(s \mid \boldsymbol{\pi}^{\text{eq}}_{-i}, \pi_{LS}\right)  & \geq V^{\pi_i}_i\left(s \mid  \boldsymbol{\pi}^{\text{eq}}_{-i}, \pi_{LS}\right) \\ 
\end{aligned}
\]
where $\boldsymbol{\pi}^{\text{eq}}_{-i}$ denotes the equilibrium strategy of agents other than $i$. Assuming finite states and actions and discount factor $\gamma \in [0,1)$ for the agents, an MPE always exists \citep{MPEexistence}\citep{Fink1964EquilibriumIA}. 
\\
\textbf{Naive follower responses.} We focus on settings with naive follower responses that are myopic and common knowledge for any corresponding leader action. Formally given a leader $\pi_l$ and its action $a_l$ at a state $s$ of a game, the best response of the followers induces a Nash equilibrium at that state, such that their instantaneous rewards are maximized. i.e.,
\[
r_i \left(s, (a_l, \boldsymbol{a}_{-l}^\text{eq}) \mid {\pi_l, \boldsymbol{\pi}_{-l}^\text{eq}}\right) \geq r_i \left(s, (a_l, \boldsymbol{a}_{-l}) \mid {\pi_l, \boldsymbol{\pi}_{-l}}\right) \forall i \in \mathcal{I} \backslash l,
\]  
and the best response $\boldsymbol{\pi}_{-l}^\text{eq}$ is common knowledge; that is, $\boldsymbol{\pi}_{-l}^\text{eq}$ is known by all agents including the leader, and the followers (assuming rationality) can directly use it to take actions. 

Such settings subsume a special form of Markov games called leader-controller Markov games, where the transition probabilities at any state solely depend on the current leader's action. Such a setting is also well-motivated; one can consider the example of a rotating government council, where the leader is the current council president, who sets the policies, and the followers are the council members. The council members' responses naively depend on the policy set by the current leader. In this case, the future state (policies) is determined by the current state (policies) and the current leader (council president). Formally, it holds that $s^{t+1} \sim \mathcal{P}\left(s^{t+1} \mid s^t, {a}_l^{{t}}\right)$ with $a_l$ being the action of the selected leader at time $t$.

\subsection{Leader selection and fairness}
As discussed in Section \ref{section:Introduction}, the nature of the traditional Stackelberg game can lead to unfair outcomes if the returns are biased towards the leader or the follower. In the case of a Stackelberg game with dynamic leaders,  the leader selection strategy $\pi_{LS}$ can considerably affect fairness.\\
\textbf{Example 1.}
\textit{Consider the normal form game \textit{"battle of the sexes"} (Table \ref{Table: BoS}). There are two players, X (the row player) and Y (the column player). X prefers to go to a Movie, while Y prefers to go to a Ballet. This game has two pure Nash equilibria: going to a Movie or a Ballet. There is only one Stackelberg equilibrium per leader. If X is the leader and commits to going to a Movie, Y's best response is also to go to a Movie. In doing so, X receives a payoff of 2, while Y receives a payoff of 1. Now consider the Markov game, where the same game is repeated several times, and the players must make the same decision repeatedly. If only X remains the leader, it results in a disproportionate cumulative payoff to X compared to Y. If X and Y alternate as the leaders, the minimum cumulative payoff among the two players will be maximized. In this case, alternating the leaders each round is an optimal leader selection strategy with respect to the minimum welfare fairness 
measure. }

That way, the leader selection strategy affects the agents' payoffs. We introduce fair mediators and propose a generic framework that captures this intuition to improve fairness among agents. 
\begin{table}[]
\centering
\caption{"Battle of the Sexes" game.}
 \vspace{5pt}
\renewcommand{\arraystretch}{1.5}
\begin{tabular}{|c|c|c|}
\hline
       & \ Movie \ & \ Ballet \ \\ \hline
\ Movie \ &  2,1   & 0,0    \\ \hline
\ Ballet \ & 0,0   & 1,2    \\ \hline
\end{tabular}
\label{Table: BoS}
\end{table}
\section{Using Mediators For Leader Selection}

Markov mediators have recently been introduced in the literature \citep{mediatedRL}. These mediators act as additional independent RL agents with their own goals and rewards to maximize. We define Markov mediators for fair leader selection in Markov Stackelberg games with dynamic leaders.

Let the tuple $\mathcal{M} =  \langle \mathcal{S}_\rho, \mathcal{A}_\rho, r_\rho, \phi\rangle$ represent a fair Markov mediator. For fully observable environments, if the mediator does not encode any extra information into its observations, its state set $\mathcal{S}_\rho$ is the same as the global environmental state set $\mathcal{S}$. $\mathcal{A}_\rho$ is the action space of the mediator and is equivalent to $\mathcal{I}$, the set of agents in the Markov game. Besides, $\boldsymbol{r}_\rho: \mathcal{S}_\rho \times \mathcal{A}_\rho \rightarrow \boldsymbol{R}$ is the mediator's reward function, with vector-valued rewards $\boldsymbol{R} \in \mathbb{R}^N$. $\phi$ is a fairness measure. At time step $t$, the mediator selects a leader at state $s_\rho^t \in \mathcal{S}_\rho$ based on its policy $\pi_\rho: \mathcal{S}_\rho \rightarrow \Omega\left(\mathcal{A}_\rho\right)$, the agents take actions in Stackelberg order, and the mediator receives feedback as a reward vector from the agents $\boldsymbol{r}_\rho^t= \boldsymbol{r}^t = \left(r^t_1, \ldots, r^t_N\right) \in \mathcal{\boldsymbol{R}}$.
Next, we define the objective for learning the mediator policy $\pi_\rho$ using $\phi$. 
\subsection{Mediator objective}
In an $N$-player Markov Stackelberg game with dynamic leaders, determine a fairness-optimal mediator policy $\pi_\rho^{\mathcal{F}^*}$ that induces an MPE where the agents' joint policy corresponds to an optimally fair joint policy, i.e.,
\begin{equation}
\boldsymbol{\pi}^{\text{eq}} \mid \pi_\rho^{\mathcal{F}^*} = \argmax_{\boldsymbol{\pi}} \phi ({J}_1( \boldsymbol{\pi}),...,{J}_N(\boldsymbol{\pi})),
\end{equation}
where $\phi$ is the pre-defined fairness measure. Besides, ${J}_k(\boldsymbol{\pi}) = 
d_0V_k^{\boldsymbol{\pi}}$ is the expected returns for agent $k$ given a joint policy $\boldsymbol{\pi}$ and the initial state distribution $d_0$.

Some fairness measures popular in the MARL literature include minimum welfare, GGF, and Nash social welfare. We redefine these in the supplementary material (Section \ref{section:fairness_measures}). We use minimum welfare throughout our experiments, but it can be easily replaced with other fairness measures. \\
\textbf{Mediated Stackelberg game.} We define a Mediated Stackelberg game as a Markov Stackelberg game with dynamic leaders in which a mediator performs the leader selection process.
\section{Proposed Solution}
In the following section, we first consider the learning setting and propose an RL approach using Q-learning to find the optimal policies for agents and the mediator. We then consider a simplified setting with full information, where dynamic programming algorithms \citep{Sutton1998} are applicable.\\
\textbf{Sequential learning.}
\color{black}
Recently, in decentralized settings, sequential learning or \textit{turn-by-turn} learning has been shown to mitigate the issue of non-stationarity \citep{alternateQLearning}\citep{multiagentvalueiteration} among independent learners and also provide theoretical guarantees for convergence. Motivated by this, we allow sequential updates between the agents and the mediator and assume the central entity, consisting of the mediator, also coordinates these updates. However, as we show empirically in the supplementary material (Section \ref{section:sim_learning}), this is not strictly necessary, and even if all the agents and the mediator learn simultaneously, they usually converge to similar solutions.

In the sequential form of learning, there are multiple rounds of updates among the learners, where in any $k^{th}$ round, each agent $i\in \mathcal{I}$ and the mediator update their policy in a turn-by-turn order. This can be represented as $\pi_1^{*,k} \rightarrow \pi_2^{*,k} ... \rightarrow \pi_N^{*,k} \rightarrow \pi_\rho^{*,k}$, the optimal policies obtained after learning updates at round $k$. Ideally, each learner performs updates until convergence to the optimal policy at each turn. However, running a limited but sufficient number of updates at each turn can also lead to desired solutions such as the Nash equilibrium \citep{alternateQLearning}.

\subsection{Q-Learning framework}
We first define modified Q-functions for the agents' leader policies, which are generic and can be used for any Markov Stackelberg game with a corresponding leader selection process. We then define the Q-function for the mediator, taking into account the factors required to maximize fairness. We formally define these Q-functions as fixed points of contraction operators in Sections \ref{section:agentsQ} and \ref{section:mediatorQ} of the supplementary material. We call our learning-based framework with the mediator integrated for leader selection as \textit{Joint Agents-Mediator Q-learning framework}, and abbreviate it as \textbf{JAM-QL}.  

\subsubsection{Learning leader policies.}
Consider an agent $i$'s MDP defined by a mediator policy $\pi_\rho$ and other agents' policy $\boldsymbol{\pi}_{-i}$. The optimal $Q$-function at state $s^t$ for its leader policy depends on the expected duration $k$ following  $s^t$ during which $i$ acts as a follower, as well as the expected rewards accumulated during that period. This effect can be defined by the $k$-step temporal difference and by applying the Bellman optimality operator
\begin{equation}
Q_{i}^*(s^t, a_i)= {\hat{r}}^t_{i} + \gamma^k \mathbb{E}_{s^{t+k} \sim \mathcal{P}} \max _{a'_i} Q_{i}^*(s^{t+k}, a'_i ),
\label{eqn:agentMDP}
\end{equation}
$$
\text{where} \quad
{\hat{r}_{i}}^t = r_{i}^t + \sum_{t'=t+1}^{t+k-1}\gamma^{t'-t}r_{i}^{t'}.
$$

With $k=1$, agent $i$ only needs to consider the immediate rewards $r_{i}^t$ due to its leader action $a_i$ and the naive follower responses of other agents at $s^t$. With  $k>1$, it naively responds to other leaders in the succeeding $k$ states and therefore is not trained. Not getting selected as the leader effectively transitions the agent's leader policy from  $s^t$ directly to $s^{t+k}$ and in the process yields the additional discounted reward of $\sum_{t'=t+1}^{t+k-1}\gamma^{t'-t}r_{i}^{t'}$.
\subsubsection{Learning mediator's policy.}
\label{section:med_policy}
Consider the mediator's MDP defined by the agents' policies $\boldsymbol{\pi}$. The objective of a fair mediator is to maximize fairness over the expected returns of the agents. To this end, we define the mediator's policy in three stages:\\
(i) \textit{Promoting fair leaders.} The mediator shall maximize the fairness in expected returns at each state through the leader selection process, i.e., the optimal Q-function is given by
\begin{equation}
    Q_\rho^*(s_\rho, a_\rho) = \max_{\pi_\rho}\phi\left(\boldsymbol{Q}_\rho^{\pi_\rho}(s_\rho, a_\rho)\right), \forall s_\rho \in \mathcal{S_\rho}, a_\rho \in \mathcal{I},
    \label{eqn:med_policy}
\end{equation}
where
$$
\boldsymbol{Q}_\rho^{\pi_\rho}(s_\rho, a_\rho) = \mathbb{E}_{\pi_\rho,  \boldsymbol{\pi}} [\sum_{t=0}^{\infty} \gamma^{t} \boldsymbol{r}^t_\rho \mid s_\rho^0 = s_\rho, a_\rho^0 = a_\rho ]
$$
represents the expected returns across the $N$ agents as estimated by the mediator starting from state $s_\rho$, taking action $a_\rho$ and then following the policy $\pi_\rho$, with the corresponding scalar values given by the fairness measure $\phi$. For example, if $\phi$ is the minimum welfare function, ${Q}_{\rho}^{\pi_\rho} = \phi\left(\boldsymbol{Q}_{\rho}^{\pi_\rho}\right) = \min \left(\boldsymbol{Q}_{\rho}^{\pi_\rho}\right)$. Note that all quantities denoted by boldface here are vectors in $\mathbb{R}^{N}$.

The optimal mediator policy $\pi_\rho^*$ can be derived from $Q_\rho^*$ and represents the policy that selects leaders whose actions result in the maximum expected fairness at every stage of the Markov game. \\
(ii) \textit{Rewarding historical performance.} In non-cooperative contexts, agents are inherently selfish and prefer to maximize their rewards. Without further incentive, the mediator would simply select leaders whose selfish actions lead to \textit{less unfair} outcomes compared to alternatives; however, it is more beneficial to incentivize leaders to take fairer actions. To that end, we add the agents' historical performance as part of the mediator state information, based on which the agents can be rewarded with leadership again. This can have a major effect on some games, such as those with the first-mover advantage, as we show in the next section. 

We do this by adding a history model $\mathcal{H}$ that keeps track of the historical rewards gained by the agents and adds it to the state information. Formally, given state-action history $h^t = \{ s^1,\boldsymbol{a}^1,s^2, \boldsymbol{a}^2..s^{t-1},\boldsymbol{a}^{t-1} \}$ at time step $t$, the history model outputs a reward vector $\boldsymbol{s}_{r}$ where $\boldsymbol{s}_{r}= \sum_{t'=1}^{t-1} \boldsymbol{r}^{t'}$ and $\boldsymbol{r}^{t'} =\left(r^{t'}_1, \ldots, r^{t'}_N\right)$ and adds it to the current environmental state $s^t$. 
The state space of the mediator $\mathcal{S}_\rho$ now includes the history-reward space $\mathcal{S}_r$ where any mediator state is given by $s_\rho =  \langle s, \boldsymbol{s}_r\rangle$, preserving the Markov property \citep{historyBasedRL}. The reward function changes to $\tilde{\boldsymbol{r}}_\rho =\boldsymbol{r}_\rho + \boldsymbol{s}_r$.

The Bellman optimality equation for the multi-dimensional Q-function in \eqref{eqn:med_policy} can then be written as
\begin{equation}
\boldsymbol{Q}_\rho^{*}(s_\rho,  a_\rho)= \tilde{\boldsymbol{r}}_\rho  + \gamma \mathbb{E}_{s'_\rho \sim \mathcal{P},\mathcal{H}} \max _{a'_\rho | \phi} \boldsymbol{Q}_\rho^*(s'_\rho, a'_\rho ),
\label{eqn:mediatorMDP}
\end{equation}
where ${a'_\rho | \phi} =\operatorname{argmax}_{a'_\rho \in \mathcal{I}} \operatorname{\phi}\left( \boldsymbol{Q}_\rho^{*} (s'_\rho, a'_\rho)\right)$ with $\phi$ being the fairness measure. 
\\
(iii) \textit{Additional end-game incentive}.
With historical performance information, the mediator can incentivize leaders to take fair actions in all states except terminal ones. This can be an issue for episodic games where agents understand which states are terminal. We study this issue empirically in the supplementary material (Section \ref{section:end_game_effects}), showing that if the mediator uses only the historical information to select leaders, their actions converge to fair ones in all but the terminal states. To mitigate this end-game effect, we add an \textit{end-of-game stage} to episodic games where the mediator can threaten to perform zero-sum rewards transfer $\boldsymbol{\theta}$ among agents based on their behavior at the terminal states. This intuition is similar to allowing rewards to be transferred among agents or through a principal using formal contracts \citep{zerosumcontracts}\citep{principalagentreinforcementlearning} at any stage of the Markov game to resolve social dilemmas. In contrast, we use it only at the end of episodes as a separate stage from the regular Markov game. From the agents' perspectives, this can be considered a modification of rewards at any terminal state $s^T$. i.e.,
\[\boldsymbol{r}'(s^T, \boldsymbol{a}) = \boldsymbol{r}(s^T, \boldsymbol{a}) + \boldsymbol{\theta}(s^T,\boldsymbol{a}),\]
where $T$ is the time period of an episode, $\boldsymbol{\theta} = (\theta_1, \theta_2....\theta_N) \ \text{and} \ \sum_{i=1}^{N} \theta_i = 0, \text{with} $ $\theta_i \in [-R_{\max}, R_{\max}]$ where $R_{\max}$ is the maximum reward difference between any two agents at terminal state $s^T$. In addition, we integrate the mediator with the option to perform a minimum or null transfer $\boldsymbol{\theta} = (0, 0...0)$ in the case of ideal leader performance, which can motivate leaders to take fair actions to prevent the reward loss in terminal states.
\subsection{Full information setting}
In this section, we assume the agents and mediator have complete knowledge about the environment dynamics (transition probabilities and reward function) and perform value iteration using the defined Q-functions. Note that this would entail replacing the sample-based updates with model-based updates in \eqref{eqn:agentMDP} and \eqref{eqn:mediatorMDP} \cite{Sutton1998}\cite{RLTheorybook}. Then, the optimal value functions of the agents and the mediator are given by 
$
V_i^*(s)=\max _{a_i} Q_i^*(s, a_i) $ for all $i \in \mathcal{I},
\ \text{and} \
{V}_\rho^*(s_\rho) =\max _{a_\rho|\phi} \phi\left(\boldsymbol{Q}_\rho^*(s_\rho, a_\rho)\right).
$
We provide convergence guarantees for full-information settings to optimally fair agents' policies and the optimal mediator policy under certain conditions. First, we define a common variant of Stackelberg games in the Markov setting.\\
\textbf{First mover-advantage Markov games.} 
\textit{A Markov Stackelberg game in which being a leader at each stage (rather than a follower) of the Markov game leads to better-expected returns for all agents, even at the expense of immediate sub-optimal returns. }\\
\textbf{Proposition.} \textit{Under the assumption that every agent has a unique leader action at each state $s \in \mathcal{S}$ that maximizes its expected selection as the leader by the mediator in future states, the policy sequence of agents in the full information setting for first-mover advantage games converges to a Markov perfect equilibrium, under a fairness-optimal mediator policy $\pi_\rho^{\mathcal{F}^*}$. }\\

We prove the proposition in Section \ref{section:proposition_proof} of the supplementary material. The proposition implies that the agents converge to an optimally fair joint policy $\boldsymbol{\bar\pi}^*$ at MPE under $\pi_\rho^{\mathcal{F}^*}$,  i.e.,  

$$
\boldsymbol{\bar\pi}^* \mid \pi^{\mathcal{F}^*}_{\rho} = \boldsymbol{\pi}^{\text{eq}} \mid \pi^{\mathcal{F}^*}_{\rho} = \argmax_{\boldsymbol{\pi}} \phi ({J}_1( \boldsymbol{\pi}),...,{J}_N(\boldsymbol{\pi})).
$$
Hence, our mediator-integrated framework can lead to an optimal level of fairness in expected returns among agents.

\begin{table}[t]
\centering
\caption{Prisoner's Dilemma}
\begin{tabular}{cc|cc}
\hline
\multicolumn{2}{c|}{Actions}             & \multicolumn{2}{c}{Payoffs}  \rule{0pt}{10pt}           \\ \hline
\multicolumn{1}{c|}{Leader}   & Follower & \multicolumn{1}{c|}{Leader} & Follower \rule{0pt}{10pt}  \\ \hline
\multicolumn{1}{c|}{cooperate} & cooperate   & \multicolumn{1}{c|}{2}      & \ 1       \rule{0pt}{10pt}  \\
\multicolumn{1}{c|}{defect}   & defect & \multicolumn{1}{c|}{3}  & -2        \\
\hline
\end{tabular}
\label{table:envPD}
\end{table}

\begin{table}[t]
\caption{Chicken}
\centering
\begin{tabular}{cc|cc}
\hline
\multicolumn{2}{c|}{Actions}             & \multicolumn{2}{c}{Payoffs}  \rule{0pt}{10pt}           \\ \hline
\multicolumn{1}{c|}{Leader}   & Follower & \multicolumn{1}{c|}{Leader} & Follower \rule{0pt}{10pt}  \\ \hline
\multicolumn{1}{c|}{straight} & swerve   & \multicolumn{1}{c|}{7}      & \ 2       \rule{0pt}{10pt}  \\
\multicolumn{1}{c|}{swerve}   & straight & \multicolumn{1}{c|}{2}      & 7        \\
\multicolumn{1}{c|}{brake}    & brake    & \multicolumn{1}{c|}{6}      & 6        \\ \hline
\end{tabular}
\label{table:envChicken}
\end{table}

\begin{figure}[t]
    \centering
      \includegraphics[width=1.0\linewidth]{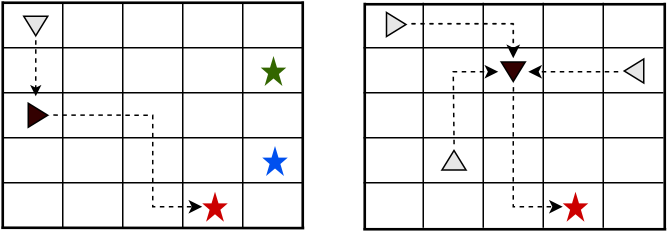}
       \caption{Resource collection environment consisting of two players (left) and four players (right). A leader (black colored) is selected, who decides which resource to collect next, and the other agents follow the leader (gray colored). Each leader prefers to collect specific types of resources (red, blue, or green stars), but they need the help of their followers.}   \label{fig:matrix_experiments}
    \vspace{15pt}
\end{figure}

\section{Experimental Setup}
We test on different classes of games. We first test on repeated matrix games, followed by higher-dimensional resource collection games. For the simpler iterated matrix games, we use independent tabular Q-learning. We utilize function approximation with deep Q-networks (DQN) for the more complex resource collection environments for agents and the mediator. We add another DQN network to learn the history model of the mediator. The detailed hyperparameters for these networks are provided in the supplementary material (Section \ref{section:hyperparameters}). In all experiments, we use the minimum welfare fairness measure for evaluation. 

\subsection{Iterated matrix games.} 
\textit{Prisoner’s Dilemma (PD). } We study the leader-controller version of the Prisoner's Dilemma (Table \ref{table:envPD}), similar to the one studied by Novak and Sigmund \cite{alternatingPD}. In this version, every agent has two actions: \textit{cooperate} or \textit{defect}. If the selected leader \textit{cooperates} (\textit{defects}), the naive response of a follower is also to \textit{cooperate} (\textit{defect}). The leader always earns a higher payoff but has more to gain with the \textit{defect} action.\\
\textit{Chicken.} Similar to Prisoner's Dilemma, we study the leader-controller version of the Chicken game (Table \ref{table:envChicken}). In this case, if the leader takes the \textit{straight} (\textit{swerve}) action, a follower's naive response is to take the \textit{swerve} (\textit{straight}) action. However, the fair outcome is always taking a third action: \textit{brake}.

For these, we consider games with 2 and 4 players. We play each game for four steps per episode, with leader selection occurring at every step. 

\begin{figure}[t]
    \centering
      \includegraphics[width=1.0\linewidth]{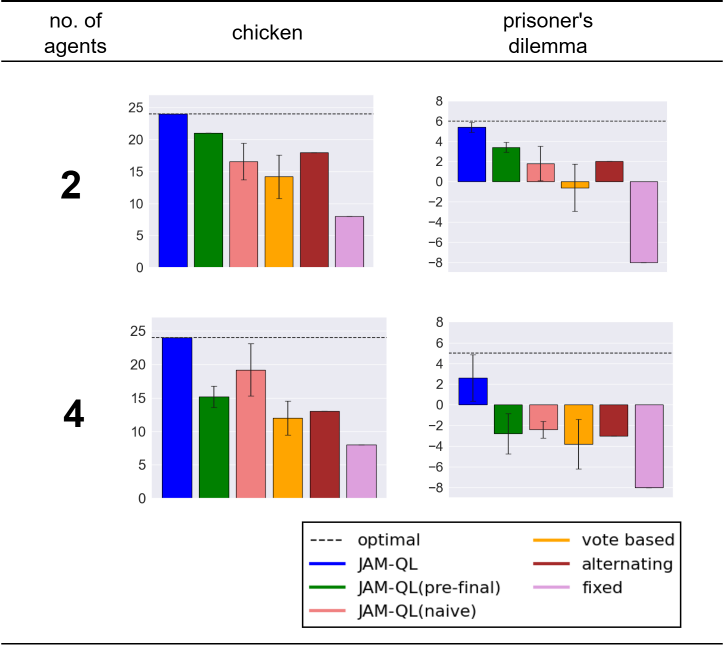}
       \caption{Experimental results for the iterated matrix games. Every plot displays the minimum welfare per episode at the end of training (200k episodes for all runs). Cells vary in the number of agents present (2, 4) and environments (chicken, prisoner's dilemma), with each cell comparing the different models (ours and baselines). Error bars denote one standard deviation over five independent runs.}
    \label{fig:matrix_experiments}
    \vspace{15pt}
\end{figure}
\subsection{Resource collection}
Resource collection is a mixed cooperative-competitive environment, where the goal is to collect certain types of resources \cite{resource_collect}. We study a leader-follower form of the environment, in which, at any stage, the leader decides which resource to collect while the followers follow the leader's direction. Leader selection occurs at game states where the following resource to collect must be determined, i.e., at the beginning of the episode or after any resource has been collected. 

Again, we consider two-player and four-player versions of the environment. We first define the two-player version. There are two agents, $A$ and $B$, and three types of resources - \textit{red, blue}, and \textit{green}. Agent $A$ ($B$) has the preference and skill to lead and collect \textit{red} (\textit{blue}) resources, while both agents possess the skill to lead and collect \textit{green} resources. However, to collect any resource, a leader needs the help of at least one follower. Collecting \textit{red} or \textit{blue} resources results in unfair returns, favoring the agent with a preference for it. Collecting \textit{green} resources yields fair and equal returns. 

We consider two different settings:
\begin{enumerate}
    \item RC-1: Only two types of resources are present - \textit{red}/\textit{blue} (randomized) and \textit{green}, and a maximum of two resources can be collected. As long as the leader with a preference for the unfair resource is not selected, fairness can be optimized. 
    \item RC-2: All types of resources are present - \textit{red}, \textit{blue}, and \textit{green}, and a maximum of two resources can be collected. In this case, both agents prefer collecting unfair resources; therefore, fairness can only be optimal if additional incentives exist to collect fair resources.
\end{enumerate}
The agents and the resources are initialized at random locations within a grid. Any agent, whether as a leader or follower, can take four actions at any time: "move forward," "move backward," "turn left," and "turn right." The leader can also take an additional action: "collect." Collecting an unfair resource yields a reward of \{5, 1\} in favor of the leader, whereas collecting a fair resource results in rewards of \{4, 4\}. At any timestep, the follower agent receives an auxiliary reward for staying close to the leader. We set the mediator's objective to maximize the fair distribution of the rewards from the resources collected. 

For the four-player version, we extend the scenario by adding two more agents, C \& D, with the same preferences and skills as A \& B, respectively. All other environmental conditions remain the same. We include a separate description of the four-player environment in the supplementary material (Section \ref{section:RC_4P}). 

\begin{figure*}[t]
    \centering
    \begin{subfigure}[b]{0.261\textwidth}
        \includegraphics[width=\textwidth]{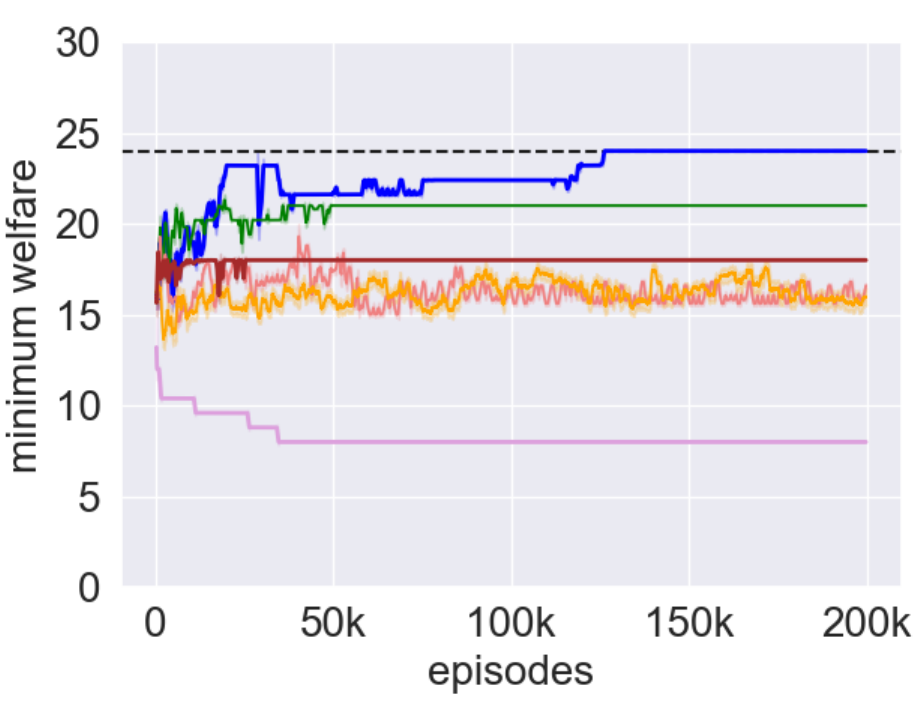}
        \caption{Chicken (2 agents)}
        \label{fig:subfig1}
    \end{subfigure}
    \begin{subfigure}[b]{0.23\textwidth}
        \includegraphics[width=\textwidth]{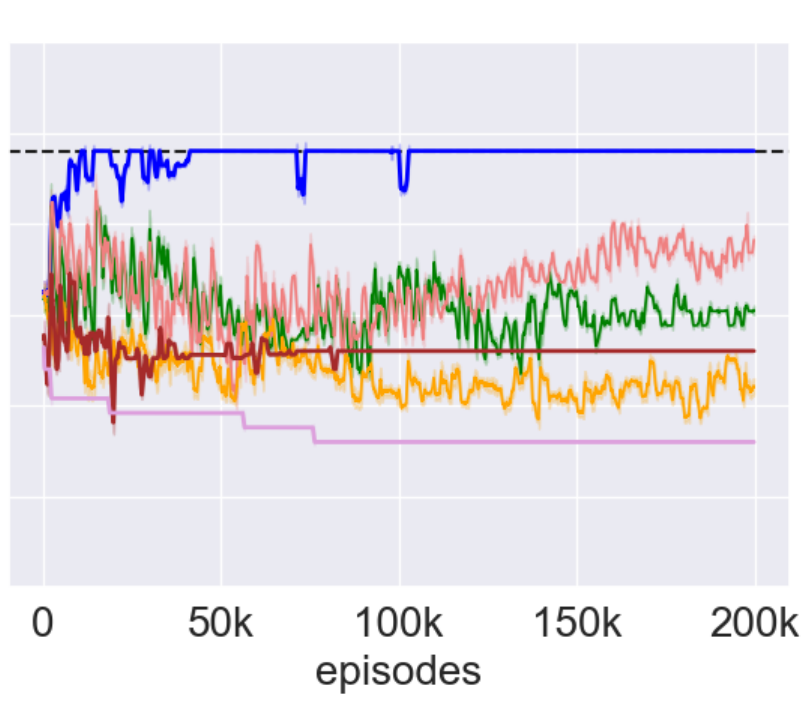}
        \caption{Chicken (4 agents)}
        \label{fig:subfig2}
    \end{subfigure}
    \begin{subfigure}[b]{0.249\textwidth}
        \includegraphics[width=\textwidth]{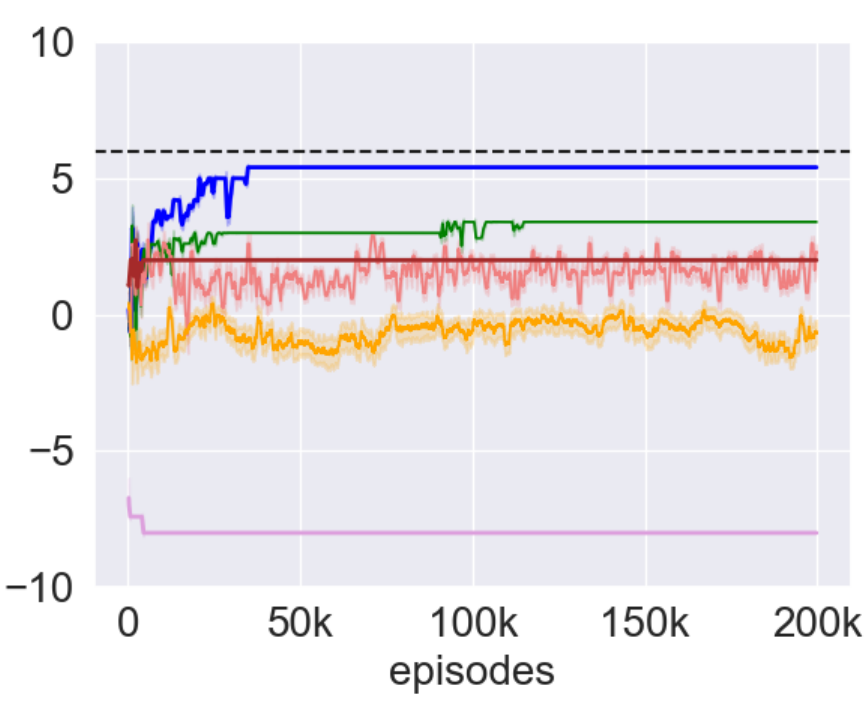}
        \caption{PD (2 agents)}
        \label{fig:subfig3}
    \end{subfigure}
    \begin{subfigure}[b]{0.23\textwidth}
        \includegraphics[width=\textwidth]{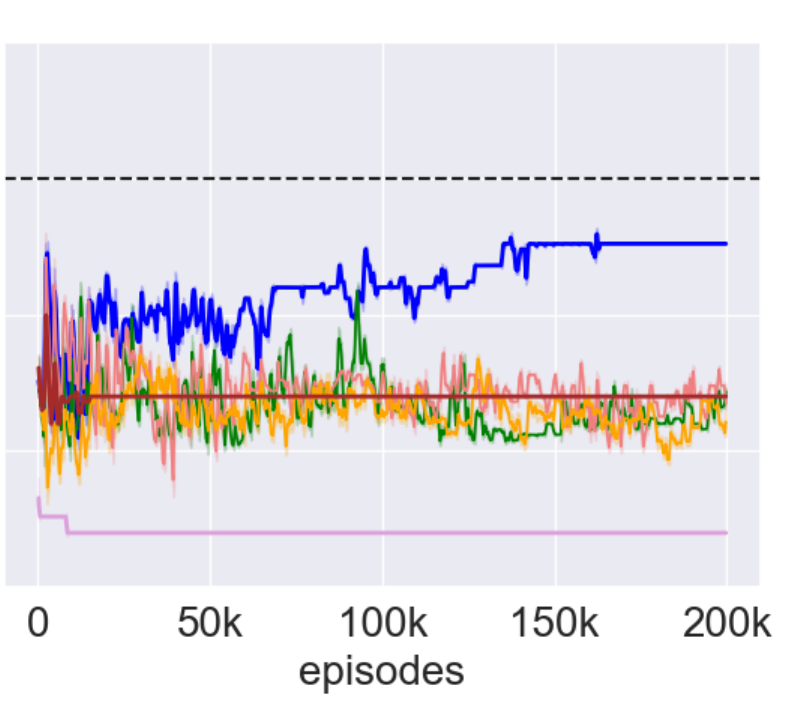}
        \caption{PD (4 agents)}
        \label{fig:subfig4}
    \end{subfigure}
    \vspace{15pt}
    \caption{Evaluation of minimum welfare through the training runs of the different models for the iterated matrix games of Chicken (a \& b) and Prisoner's Dilemma (c \& d). Color coding remains the same, as in Figure \ref{fig:matrix_experiments}. Results are averaged over five independent runs.}
    \label{fig:results_trajectory_matrix}
    \vspace{15pt}
\end{figure*}
\subsection{Baselines}
We use the following baselines for comparison with our JAM-QL framework:
\begin{enumerate}
    \item fixed: There is a fixed leader at all game stages.
    \item alternating: The agents alternate as leaders at every leader selection stage.
    \item vote-based: The agents vote for a leader at each leader selection stage before taking the next set of actions, with ties broken randomly.
    \item JAM-QL(naive): We integrate a version of the mediator that naively maximizes expected fairness at each state without providing any additional incentives for the leaders to take fair actions, i.e., without stages (ii) and (iii) in Section \ref{section:med_policy}.
    \item JAM-QL(pre-final): We integrate a version of the mediator that does not consider end-game effects,  thereby not providing the additional end-game incentive to agents, as defined in stage (iii) of Section \ref{section:med_policy}.
\end{enumerate}

\section{Results}
\begin{figure*}[t]
    \centering
    \begin{subfigure}[b]{0.235\textwidth}
        \includegraphics[width=\textwidth]{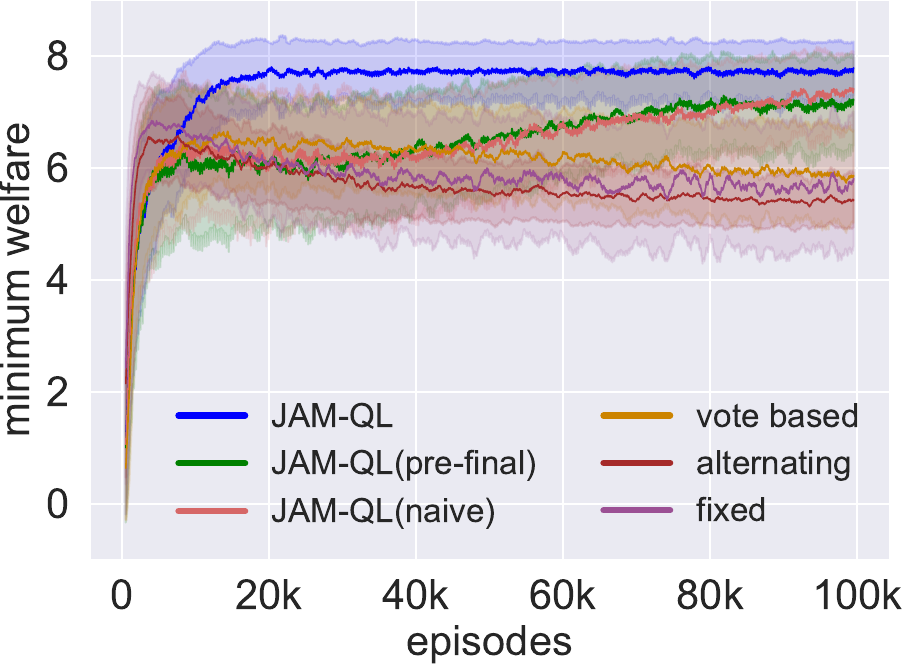}
        \caption{RC-1 (2 agents)}
        \label{fig:subfig1}
    \end{subfigure}
    \begin{subfigure}[b]{0.235\textwidth}
        \includegraphics[width=\textwidth]{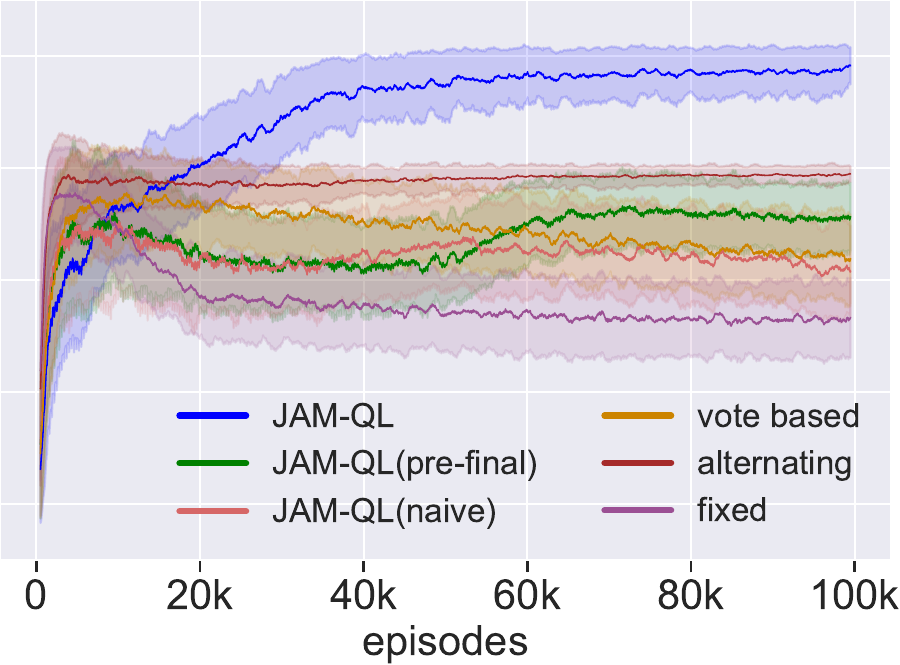}
        \caption{RC-2 (2 agents)}
        \label{fig:subfig2}
    \end{subfigure}
    \begin{subfigure}[b]{0.235\textwidth}
        \includegraphics[width=\textwidth]{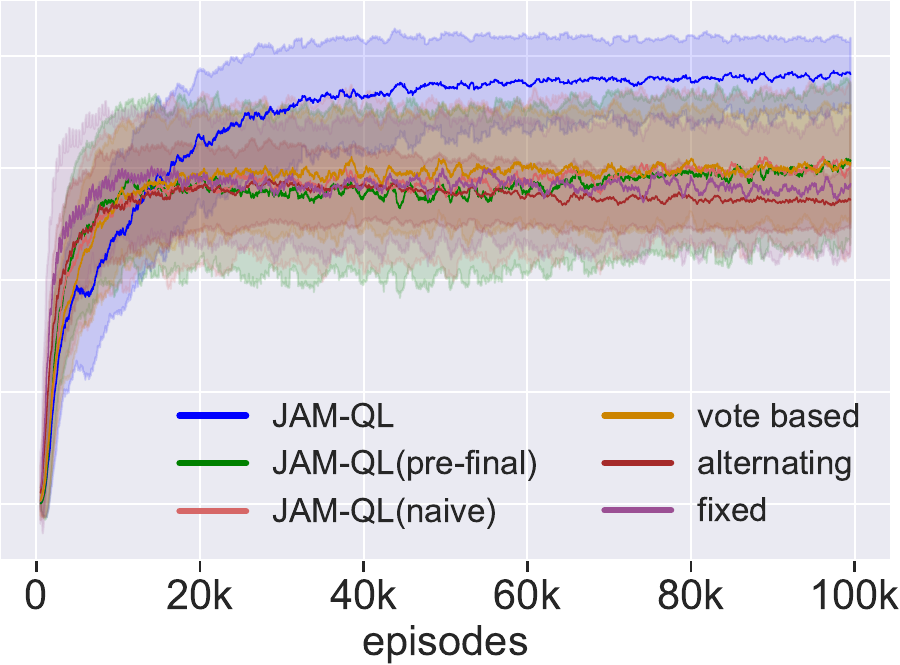}
        \caption{RC-1 (4 agents)}
        \label{fig:subfig3}
    \end{subfigure}
    \begin{subfigure}[b]{0.235\textwidth}
        \includegraphics[width=\textwidth]{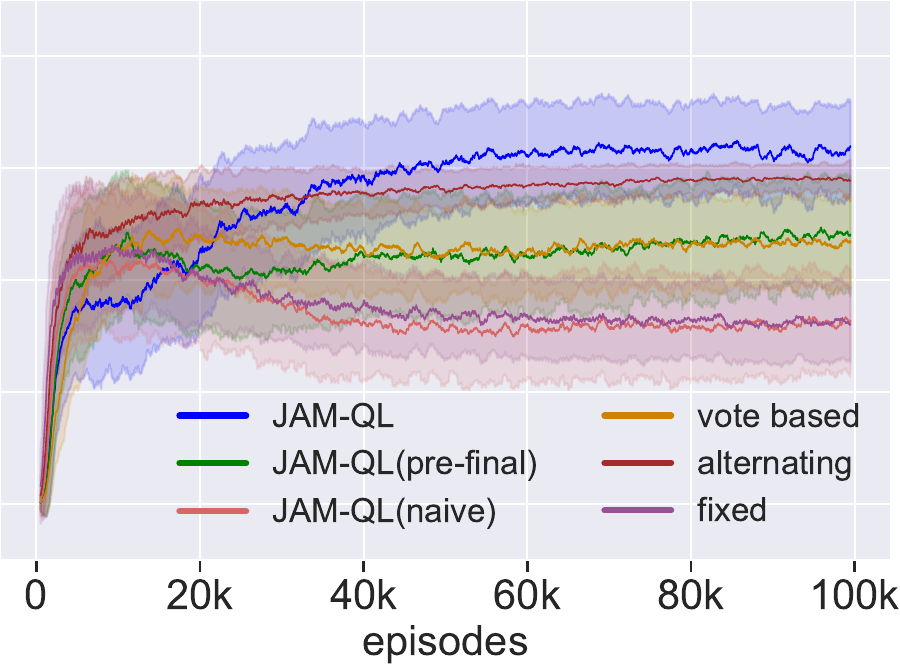}
        \caption{RC-2 (4 agents)}
        \label{fig:subfig4}
    \end{subfigure}
    \vspace{15pt}
    \caption{Minimum welfare of all approaches during the learning phase for the two-agent version (a \& b) and four-agent version (c \& d) of resource collection environments. Results are averaged over five independent training runs.}
    \label{fig:all_subfigures}
    \label{fig:results_RC}
    \vspace{15pt}
\end{figure*}

\textit{Iterated matrix games.} Figures \ref{fig:matrix_experiments} and \ref{fig:results_trajectory_matrix} illustrate the performance of JAM-QL compared to the baselines for the iterated matrix games. Across all games, JAM-QL yields higher levels of fairness compared to the baselines. It achieves the optimal level of fairness for both the 2- and 4-player versions of the chicken game and is close to this level for the prisoner's dilemma. For the chicken game, selecting and incentivizing one of the agents to take fair actions as a leader is sufficient. However, in the prisoner's dilemma, all agents have to be selected as leaders in the correct order, and all of them must have the incentive to take fair actions. Hence, the complexity increases for the mediator. Still, JAM-QL performs much better than the baselines. For the baselines, JAM-QL(pre-final) is the closest in terms of overall performance, but it suffers from possible end-game defections by the agents. Similarly, the lack of clear incentives for agents in JAM-QL(naive) results in leaders switching between fair and unfair actions, leading to high variance in overall performance. The vote-based mechanism also exhibits high variance, as selfish agents often fail to reach a common consensus. These variations are best seen in the Figure \ref{fig:results_trajectory_matrix}. Alternating agents as leaders can exhibit a natural improvement in fairness to a certain extent, as observed in the results of the chicken game. However, since it is only a rigid rule and not optimized explicitly for fairness, it does not achieve an optimal level. Finally, having a single selfish agent as the leader can yield the most unfair returns.\\
\textit{Resource collection.} We compare the performance of JAM-QL with the baselines for the resource collection environments RC-1 and RC-2 in Figure \ref{fig:results_RC}. First, we consider the two-player version in Figures \ref{fig:results_RC}(a) and \ref{fig:results_RC}(b). Fairness in RC-1 is optimal if the right leaders are selected, i.e., those with no selfish action to take. Thereby, naively maximizing for expected fairness can result in optimal fairness; no end-game effects also occur. Hence, both JAM-QL(naive) and JAM-QL(pre-final) converge to optimal fairness levels. However, JAM-QL converges faster, indicating that using our complete framework solely for selecting the optimal order of agents as leaders remains beneficial. The other baselines are not optimized for fairness, suffer from selfish agent actions, and hence cannot achieve optimal fairness even in RC-1. Furthermore, in the case of the  RC-2 environment, only JAM-QL consistently provides the necessary incentive for agents to take fair actions as leaders, similar to the iterated matrix games, resulting in superior performance compared to the baselines.

Next, we compare the performance for the four-player version in Figures \ref{fig:results_RC}(c) and \ref{fig:results_RC}(d). In contrast to the two-player version, JAM-QL(naive) and JAM-QL(pre-final) struggle to converge to optimal fairness levels and exhibit high variance in the RC-1 environment. We speculate that this occurs because some leaders can switch between collecting fair and unfair resources, confounding the mediator and making its learning process harder. The performance of JAM-QL remains stable, which confirms that having the proper incentive structure is beneficial, even for just selecting the optimal leaders. In the case of the RC-2 environment, the performance of the different models follows a similar trend to the two-player version, with JAM-QL maintaining its superior performance.

\section{Related Works}
\textbf{Stackelberg games with dynamic leaders.} The majority of works in the literature have focused on Stackelberg games with a fixed leader and follower \citep{Conitzer2006ComputingTO}\citep{brero2022learningmitigateaicollusion}\citep{gerstgrasser2023oracles}. In this setting, the leader learns a utility-maximizing strategy to commit to, and the follower best responds to this strategy. This model has been extensively studied, especially for security applications \citep{security1}\citep{security2}.
Settings where the role of leader and follower can be interchanged are widely under-explored in the literature. However, there are many real-world applications where this is possible. Examples include changing majority and minority shareholders in a corporation, different players acting as leaders in each round of the contract bridge card game \citep{NIE2008}, and biological situations such as switching leaders among bats or baboons for food or mating purposes \citep{alternatingPD}\citep{PDwithoutSynchrony}. Few works that have explored this direction include settings where agents alternate as leaders \citep{alternatingPD}\citep{PDwithoutSynchrony}, or change leaders among themselves through agreements or voting \citep {BeALeaderOrFollower}\citep{traffic_routes}\citep{who_leads}\citep{vote_based}. In contrast with these, we formalize the problem of having dynamic leaders and integrate mediators to perform the leader selection process.\\
\textbf{Fairness in sequential MARL.}
Previous works on fairness in multi-agent reinforcement learning (MARL) have mostly considered the setting of \textit{equal}, \textit{simultaneously acting }agents \cite{jiang2019learning}\cite{zimmer2021learning}\cite{mandal2023socially}\cite{ju2023achieving}. Moreover, these works assume that the agents are fully cooperative, allowing agents to share rewards arbitrarily \cite{jiang2019learning} or expected returns \cite{zimmer2021learning}. In contrast, our work assumes that agents act in their self-interest, and their emergent behavior in the mediator's presence drives them to take fair actions. Moreover, agents can only share private information with trusted central mediators. Further, we consider fairness in \textit{sequentially acting} agents with a leader-follower hierarchy within the context of Stackelberg games. Few works have a close relation to fairness in sequential MARL. Guo et al. \cite{traffic_routes} optimize for traffic congestion using multi-agent Q-learning, in which agents play a Stackelberg game, with the agent having the largest traffic queue automatically getting promoted as the leader. They optimize for system efficiency and do not explicitly consider fairness. Separate from the MARL literature, Fuzhou et al. \cite{auction} introduced the concept of fair Stackelberg equilibrium in competitive auctions, at which two
risk-seeking insiders have an equal chance to be a leader or follower at each auction stage. Our work considers a generalized version of this concept since all agents have an equal chance to be the leader at any stage through the leader selection process of the mediator.\\
\textbf{Mechanism design.} 
Our work fits within the wide literature of mechanism design \cite{mechanism_design1}\cite{mechanism_design2}\cite{mechanism_design3}\cite{mechanism_design4}. Mechanism design studies where and how to design
rules or institutional frameworks to align individual agents' incentives so that a socially desirable outcome can be achieved \cite{openproblemscooperativeai}. The central mediators we define can be considered as one such institutional framework. A closely related set of works within the scope of mechanism design use contracts from a central entity to guide agents' policies \cite{kimplementation}\cite{principalagentreinforcementlearning}. Such formal contracts can be used to provide an extrinsic form of incentive; in contrast, we develop an intrinsic form of incentive mechanism through the leader selection process of the mediators.\\
\textbf{Emergent prosocial behavior.} The emergence of prosocial behavior in human and animal societies, as well as in systems of artificial agents, has been a central research topic for many years \cite{prosocial1}\cite{prosocial2}\cite{prosocial3}\cite{openproblemscooperativeai}. Investigating strategies and 
techniques required for such emergence is fundamental for the
development of artificial agents capable of effectively collaborating with other agents and humans. Naive MARL approaches often lead to defective agent behavior due to individual selfishness \cite{LoLA}\cite{prosocial4}. To mitigate this, prior works have introduced strategies such as opponent modeling,  appropriate reward formulations, or peer incentivization \cite{LoLA}\cite{jaques2019socialinfluenceintrinsicmotivation}\cite{gifting}. Our work explores an alternative direction and shows that integrating central mediators with the right inbuilt incentives can also cause prosocial behaviors to emerge.

\section{Conclusion}
In this work, we introduce mediators in the context of Stackelberg games, where leaders can change dynamically. After formally defining the problem, we propose an RL-based framework with agents and mediators. By incorporating fairness into the mediator's definition, we demonstrate how it can lead to agents emerging as fair leaders and theoretically prove their convergence to optimal fair policies under certain assumptions. We also offer a deep RL implementation and empirically validate the benefits of our approach. Our research, particularly the integration of Markov mediators into Stackelberg games, raises numerous exciting follow-up questions. Firstly, we considered settings with leadership advantages. However, there are scenarios where it is advantageous to be followers instead, for example, if the information disclosed by
the leader can be exploited by followers, as in insider trading in a financial market model \cite{auction}. Investigating the complete dynamic of both leader and follower advantages is a compelling challenge. Secondly, we investigated scenarios with single leaders and multiple followers; adapting to multiple leaders presents a separate challenge. Finally, our proposed framework is based on Stackelberg games with naive follower responses; adapting it further without this restriction can extend its applicability.   

Overall, we anticipate that this research will encourage the RL community to further explore along two directions: (i) using central institutions or entities, such as mediators, to promote more equitable and benevolent behavior by self-serving agents, and (ii) ingraining the right incentives within the frameworks of multi-agent systems, due to which such behaviors can naturally arise. 

\begin{ack}
This research was supported by the German Research Foundation (DFG) under Grant MA7111/6-1 and MA7111/7-1. The authors gratefully acknowledge the Paderborn Center for Parallel Computing (PC2) for providing the computational resources used in this work.
\end{ack}


\newpage
\bibliography{mybibfile}

\newpage
\onecolumn
\begin{center}
    \Huge \textbf{Supplementary Material}
\end{center}
\section{Fairness Measures}
\label{section:fairness_measures}
We define fairness in multi-agent learning by considering the expected returns $J_i\left(\boldsymbol{\pi}\right)$ of each agent $i \in \mathcal{I} $ for a given joint policy $\boldsymbol{\pi}$. In particular, a fairness measure $\phi$ should map the set of expected returns of all agents to a real number, i.e., $\phi: \mathbb{R}^N \rightarrow \mathbb{R} $. To optimize for fairness, we focus on optimizing equity and efficiency; we want the agents to converge to a joint policy $\boldsymbol{\pi}^*$ that maximizes  $\phi({J}_1\left(\boldsymbol{\pi}\right),...,{J}_N\left(\boldsymbol{\pi}\right))$. Here are some popular fairness measures that can be used as $\phi$. \\
\textbf{Minimum welfare.} This fairness measure is used to maximize the minimum expected returns across the $N$ agents,

\[
 \max_{\boldsymbol{\pi}} \min _{i \in \mathcal{I}} {J}_i\left(\boldsymbol{\pi}\right).
\]
\textbf{Generalized Gini social welfare  function (GGF).} This notion of fairness generalizes max-min fairness and encapsulates properties of \textit{efficiency, impartiality, and equity} \cite{zimmer2021learning}. We are given a vector of weights $w \in \mathbb{R}^N$ so that $w_i \geq 0$ for each $i, \sum_i w_i=1$, and $w_1 \geq w_2 \geq \ldots \geq w_N$. Given a policy $\boldsymbol{\pi}$, let $i_1, \ldots, i_N$ be an ordering of the agents so that ${J}_{i_1}\left(\boldsymbol{\pi}\right) \leq {J}_{i_2}\left(\boldsymbol{\pi}\right) \leq \ldots \leq {J}_{i_N}\left(\boldsymbol{\pi}\right)$. Then, this fairness measure defines the following objective,
\[
\max_{\boldsymbol{\pi}} \sum_{k=1}^N w_{k} {J}_{i_k}\left(\boldsymbol{\pi}\right)  .
\]
\textbf{Nash social welfare.} This fairness measure is used to maximize the product of the expected returns of the $N$ agents,
\[
\max_{\boldsymbol{\pi}} \prod_{k=1}^N  {J}_{i}\left(\boldsymbol{\pi}\right)  .
\]

\section{Proposition Proof}
\label{section:proposition_proof}
\textbf{Proposition.} \textit{Under the assumption that every agent has a unique leader action at each state $s \in \mathcal{S}$ that maximizes its expected selection as the leader by the mediator in future states, the policy sequence of agents in the full information setting for first-mover advantage games converges to a Markov perfect equilibrium, under a fairness-optimal mediator policy $\pi_\rho^{\mathcal{F}^*}$. }\\
\textbf{Proof.} Consider the sequential updates in the JAM-VI framework, where the $(k+1)^{th}$ round of updates can be represented with the effective mediator value function at each point as,
\begin{align*}
V_{\rho}^{*,k}, \pi_\rho^{*,k} \rightarrow \{V^{*,k+1}_{1}, \pi_1^{*,k+1}\},V^{*,k}_{\rho,1} \rightarrow \{V^{*,k+1}_{2},\pi_2^{*,k+1}\},V^{*,k}_{\rho,2} \ \
..... \ \ \{V^{*,k+1}_{N},\pi_N^{*,k+1}\},V^{*,k}_{\rho,N} \rightarrow V^{*,k+1}_{\rho},\pi_\rho^{*,k+1}.
\end{align*}
To prove the proposition, we first show that the following monotonic improvement of the mediator's value function holds,  
\begin{equation}
\centering
{V}^{*,k+1}_{\rho} \ge V^{*,k}_{\rho,N} \ge  V^{*,k}_{\rho,N-1} \ ..... \ V^{*,k}_{\rho,2} \ge V^{*,k}_{\rho,1} \ge V^{*,k}_{\rho} \ge V^{*,k-1}_{\rho, N}.
\label{eqn:monotonic}
\end{equation}
We prove the relation starting from the right side. Because of the monotonic improvement of the Bellman optimality operation, we know that $V^{*,k}_{\rho} \ge V^{*,k-1}_{\rho, N}$ from the mediator's update. Next, we show the set of inequalities $ V^{*,k}_{\rho,N} \ge  V^{*,k}_{\rho,N-1} \ ... \ V^{*,k}_{\rho,2} \ge V^{*,k}_{\rho,1} \ge V^{*,k}_{\rho}$ hold. 

In a first-mover advantage game, there is an incentive for each agent to be selected as the leader at any state $s$ of the Markov game. Since the agents are indifferent to any historical reward information $\boldsymbol{s}_r \in S_r$ used by the mediator, by extension, there is an incentive to be the leader at any mediator state $s_\rho = \langle s, \boldsymbol{s}_r \rangle$.\\
Now, let us assume after $k$ round of updates, the mediator policy  $\pi_\rho^{*,k}$ selects the agent $i \in \mathcal{I}$ as leader at state $s_\rho$ using its estimated Q-function, which is based on agent $i$ taking the action $a_{i,1}$, i.e.
\begin{equation}
\pi_\rho^{*,k}(s_\rho) = i = \argmax_{a_\rho} \phi\left(\boldsymbol{Q}_{\rho}^{*,k}(s_\rho, a_\rho \mid \pi_i^{*,k}(s) = a_{i,1})\right).
\label{eqn:obs1}
\end{equation}
Then, to be selected as the leader again after $s_\rho$ (which is beneficial because of the first mover-advantage), agent $i$ has to 
take an action that leads to future states where the expected probability of the mediator to select agent $i$ as leader again is improved. For example, this leadership probability is maximized for all possible next states after $s_\rho$ if 
\begin{equation}
\mathbb{E}_{s'_\rho \sim \mathcal{P},\mathcal{H}} \phi \left( \boldsymbol{Q}_\rho^{*,k}(s'_\rho, a^{\prime}_\rho = i) \right) \ge \mathbb{E}_{s'_\rho \sim \mathcal{P},\mathcal{H}} \phi \left( \boldsymbol{Q}_\rho^{*,k}(s'_\rho, a^{\prime}_\rho = j) \right) \forall j \in \mathcal{I}/i.
\label{eqn:obs1}
\end{equation}
The same holds for all future states that can be reached from $s_\rho$. Further, due to the incentive structure of the mediator (stage 2), the expected probability of agent $i$ getting selected as the leader again improves if agent $i$ takes a fairer action, for example taking an action $a_{i,2}$ where $\phi\left(\boldsymbol{\tilde{r}}'_\rho(s_\rho, a_\rho = i \mid \pi_i(s) = a_{i,2})\right) \ge \phi\left(\boldsymbol{\tilde{r}}_\rho(s_\rho, a_\rho = i \mid \pi_i(s) = a_{i,1})\right)$, which will result in, 

\[
\begin{aligned}
{V}_\rho^{{*,k}}(s_\rho) & = \phi\Big(\boldsymbol{Q}_{\rho}^{*,k}(s_\rho, a_\rho = i) \mid \{{\boldsymbol{\pi}}^{*,k}\} \Big)\\
& = 
\phi\Big(\tilde{\boldsymbol{r}}_\rho(s_\rho, a_\rho = i \mid \pi_i^{*,k}(s) = a_{i,1})  +   \gamma \mathbb{E} \max _{a^{\prime}_\rho | \phi} \boldsymbol{Q}_\rho^{{*,k}}(s'_\rho, a^{\prime}_\rho) \mid \{{\pi_i}^{*,k}, {\boldsymbol{\pi}}_{-i}^{*,k}\} \Big)\\
& \le \phi\Big(\tilde{\boldsymbol{r}}'_\rho(s_\rho, a_\rho = i \mid \pi_i^{*,k+1}(s) = a_{i,2}) + \gamma \mathbb{E} \max _{a^{\prime}_\rho | \phi} \boldsymbol{Q}_\rho^{{*,k}}(s'_\rho, a^{\prime}_\rho) \mid \{{\pi_i}^{*,k+1}, {\boldsymbol{\pi}}_{-i}^{*,k}\} \Big)\\
& = \phi\Big(\boldsymbol{Q}_{\rho,i}^{*,k}(s_\rho, a_\rho = i) \mid \{{\pi_i}^{*,k+1}, {\boldsymbol{\pi}}_{-i}^{*,k}\} \Big)\\
& = {V}_{\rho,i}^{{*,k}}(s_\rho)
\end{aligned}
\]
where $\boldsymbol{Q}_{\rho,i}^{{*,k}}$ and ${V}_{\rho,i}^{{*,k}}$ are the effective mediator action-value and value functions after only agent $i$'s update in round $k+1$. This implies that due to the leadership incentive, the updates of agent $i$ can only improve the mediator's value function at any state until it has maximized its probability of being selected as the leader again in all future states.

However, at $s_\rho$, agent $i$ can have multiple actions that maximize its leadership probability in future states equally. For example, both actions $a_{i,1}$ and $a_{i,2}$ of agent $i$ can hold \eqref{eqn:obs1}.
In this case, agent $i$ can choose between $a_{i,1}$ or $a_{i,2}$ depending on its own interests and stay as the leader in the next state, and \eqref{eqn:monotonic} might not hold at $s_\rho$; for example, if agent $i$ switches from action $a_{i,2}$ to $a_{i,1}$. 
However, due to the uniqueness assumption, there is a unique action for agent $i$ at state $s_\rho$ (or $s$) that maximizes its probability of being selected as the leader in future states; let this be $a_{i,s^*}$. Further, from the incentive structure of the mediator, taking the fairest action at $s_\rho$ maximizes the probability of being selected as the leader again in future states. Hence, by extension, the unique action in this case is $a_{i,s^*} = a_{i,2}$ unless agent $i$ has a fairer action to take. Thus, the mediator's value function at $s_\rho$ can only improve. This applies to all states and agents' updates in the $(k+1)^{th}$ round, and thus the inequalities $  V^{*,k}_{\rho, N} \ge  V^{*,k}_{\rho, N-1} \ ... \ V^{*,k}_{\rho,2} \ge V^{*,k}_{\rho,1} \ge V^{*,k}_{\rho} $ in \eqref{eqn:monotonic} hold. Finally, $ {V}^{*,k+1}_{\rho} \ge V^{*,k}_{\rho, N} $ again holds due to the Bellman optimality operation of the mediator.

This proof can be extended to all rounds of agents and mediator updates; thus, the
mediator's value function improves monotonically and as $k \rightarrow \infty$, the mediator's value function converges to some function $V_\rho^{*,k \rightarrow \infty}$ with a corresponding mediator policy $\pi^{*,k \rightarrow \infty}_{\rho}$. At this point, each agent has maximized its probability of being selected as the leader by the mediator at all states. Because of the assumption, this happens only when all agents take the fairest action available to them at each state and converge to a set of policies $\{\bar{\pi}^*_1 ... \bar{\pi}^*_N\} $ s.t.
\[
\bar{\pi}^*_i(s) = a_{i,s^*} = \argmax_{a_i} Q^*_i (s, a_i) \mid \boldsymbol{\bar{\pi}}^*_{-i}, \ \forall i,s.
\]
Finally, at convergence, the mediator has selected the fairest leaders, and the leaders have taken the fairest action available to them, which implies that fairness is maximized. Moreover, the mediator policy is fairness-optimal, i.e. $V^{\mathcal{F}^*}_{\rho} = V_\rho^{*,k \rightarrow \infty}$ and $\pi^{\mathcal{F}^*}_{\rho} = \pi^{*,k \rightarrow \infty}_{\rho}$; with the joint policy $\bar{\boldsymbol{\pi}}^*$ being an optimally fair joint policy. Further, the agents' policies are implicitly at an MPE in the Markov game defined by $\pi^{\mathcal{F}^*}_{\rho}$, i.e.

$$
\boldsymbol{\bar\pi}^* \mid \pi^{\mathcal{F}^*}_{\rho} = \boldsymbol{\pi}^{\text{eq}} \mid \pi^{\mathcal{F}^*}_{\rho} = \argmax_{\boldsymbol{\pi}} \phi ({J}_1( \boldsymbol{\pi}),...,{J}_N(\boldsymbol{\pi})).
$$
\section{Agents' Q Function}
\label{section:agentsQ}
We redefine the Bellman optimality equation for the agents' Q function here.
\begin{equation}
Q_{i}^*(s^t, a_i)= {\hat{r}}^t_{i} + \gamma^k \mathbb{E}_{s^{t+k} \sim P} \max _{a'_i} Q_{i}^*(s^{t+k}, a'_i ), \forall i \in \mathcal{I},
\label{eqn:sup_agentMDP}
\end{equation}
$$
\text{with} \quad
{\hat{r}_{i}}^t = r_{i}^t + \sum_{t'=t+1}^{t+k-1}\gamma^{t'-t}r_{i}^{t'},
$$
where $s^t \in \mathcal{S}$, $a_i \in \mathcal{A}_i$ and $k$ is the expected follower period after $s^t$.\\
\textbf{Bellman optimality operator.} Consider each agent $i$'s optimization task. Solving it implies finding a fixed point of an operator $\mathcal{T}_{\rho'}$, which is the Bellman optimality operator in the agent $i$'s MDP defined by the policies $\pi_\rho$ and $\boldsymbol{\pi}_{-i}$.\\
\textbf{Definition 3.} Given an agent's MDP defined by the policies $\pi_\rho$ and $\boldsymbol{\pi}_{-i}$, the Bellman optimality operator $\mathcal{T}_{\rho'}$ is defined by
\begin{equation}
\left(\mathcal{T}_{\rho'} Q_i\right)(s^t, a_i)=\mathbb{E}_{s^{t+k} \sim \mathcal{P} \mid \{ \pi_\rho, \boldsymbol{\pi}_{-i} \}}\left[\left(\hat{r}_i(s^t, a_i)+\gamma^k \max _{a'_i} Q_i\left(s^{t+k}, a'_i\right)\right)\right].
\label{eqn:BellmanOptOp}
\end{equation}
Given $\gamma \in[0,1)$ and $k \in \mathbb{N}$, this operator is a contraction and admits a unique fixed point ${Q}_i^*( \pi_\rho, \boldsymbol{\pi}_{-i} )$ that satisfies:

\begin{equation}
{V}_i^*(s^t \mid \pi_\rho, \boldsymbol{\pi}_{-i})=\max _{a_i} {Q}_i^*(s^t, a_i \mid \pi_\rho, \boldsymbol{\pi}_{-i}),
\end{equation}
which we can find out using Q-learning.
\section{Mediator's Q Function}
\label{section:mediatorQ}
We redefine the Bellman optimality equation for the mediator's Q function here.
\begin{equation}
\boldsymbol{Q}_\rho^{*}(s_\rho,  a_\rho)= \tilde{\boldsymbol{r}}_\rho  + \gamma \mathbb{E}_{s'_\rho \sim \mathcal{P},\mathcal{H}} \max _{a^{\prime}_\rho | \phi} \boldsymbol{Q}_\rho^*(s'_\rho, a^{\prime}_\rho ),
\label{eqn:sup_mediatorQ}
\end{equation}

where $s_\rho =  \langle s, \boldsymbol{s}_r\rangle \in \mathcal{S}_\rho$ and $\tilde{\boldsymbol{r}}_\rho =\boldsymbol{r}_\rho + \boldsymbol{s}_r$ with $\boldsymbol{s}_r$ being the historical reward vector before state $s$ was reached. $s'_\rho =  \langle s', \boldsymbol{s}_r'\rangle \in \mathcal{S}_\rho$ is given by the state transition distribution $\mathcal{P}$ and history model $\mathcal{H}$. Besides, $a_\rho, a'_\rho \in \mathcal{A}_\rho$. All quantities denoted by boldface here are vectors in $\mathbb{R}^{N}$. Note that $\tilde{\boldsymbol{r}}_\rho = \tilde{\boldsymbol{r}}$ and $\boldsymbol{r}_\rho = \boldsymbol{r}$. 
We can then define a truncated optimal Q-function as, 
\begin{equation}
\boldsymbol{\bar{Q}}_\rho^  {*}(s,  a_\rho)= \boldsymbol{r}  + \gamma \mathbb{E}_{s'_\rho \sim \mathcal{P},\mathcal{H}} \max _{a^{\prime}_\rho | \phi} \boldsymbol{Q}_\rho^*(s'_\rho, a^{\prime}_\rho ),
\label{eqn:truncatedQ}
\end{equation}
which represents the mediator's Q-values, barring the historical rewards until $s$. Then we can train the mediator by learning the truncated Q-function and then compute the final Q-function using Equation \ref{eqn:sup_mediatorQ}.
\\
\textbf{Bellman optimality operator.} Consider the mediator's optimization task. Solving it implies finding a fixed point of an operator $\mathcal{T}_{\mathcal{I}_{i}}$, for each objective $i \in \{1,2, \ldots, N\}$ in the multi-objective equation \ref{eqn:sup_mediatorQ}, which are the Bellman optimality operators in the mediator's MDP defined by the joint agents' policies $\boldsymbol{\pi}$ or $\pi_\mathcal{I}$.\\
\textbf{Definition 1.} Given the mediator's MDP defined by the policies $\pi_\mathcal{I}$, the Bellman optimality operator $\mathcal{T}_{\mathcal{I}_{i}}$ for each objective is given by,

\begin{equation}
\left(\mathcal{T}_{\mathcal{I}_{i}} {Q_{\rho_i}}\right)((s, s_{r_i}),  a_\rho)=\mathbb{E}_{s',s'_{r_i} \sim \mathcal{P}, \mathcal{H}}\left[\left({\tilde{r}_i}((s, s_{r_i}) , a_\rho)+\gamma \max _{a_\rho^{\prime}|\phi} {Q_{\rho_i}}\left((s',s'_{r_i}), a'_\rho\right)\right)\right],
\label{eqn:BellmanOptOp}
\end{equation}
where {${\tilde{r}_i}((s, s_{r_i}), a_\rho)={r}_i(s, a_\rho)+ s_{r_i}$ with $s_{r_i}$ being the historical rewards specific to $i$; $Q_{\rho_i}$ is an element of a vector space $\mathcal{Q}_{S A}=\{S \times A \rightarrow \mathbb{R} \}$. This operator is a contraction, and the operators for all the objectives together admit a unique fixed point $\boldsymbol{Q}_{\rho}^*( \pi_\mathcal{I} )$ that satisfies:

\begin{equation}
{V}_\rho^*\left((s, \boldsymbol{s}_r) \mid \pi_\mathcal{I}\right)=\max _{a_\rho|\phi} \phi\left(\boldsymbol{Q}_\rho^*((s,\boldsymbol{s}_r), a_\rho \mid \pi_\mathcal{I}) \right),
\end{equation}
with 
$$
{\boldsymbol{Q}}_\rho^*((s,\boldsymbol{s}_r), a_\rho ) =  ({Q}_{\rho_1}^*((s,s_{r_1}), a_\rho ),..,{Q}_{\rho_i}^*((s,s_{r_i}), a_\rho ),..,{Q}_{\rho_N}^*((s,s_{r_N}), a_\rho ), 
$$
and
\begin{equation}
{Q}_{\rho_i}^*((s, s_{r_i}) , a_\rho)=\mathbb{E}\left[\left({\tilde{r}}_i((s, s_{r_i}), a_\rho)+\gamma \max _{a'_\rho|\phi} {Q}_{\rho_i}^*\left((s',s'_{r_i}), a'_\rho \right)\right)\right] \forall i \in \mathcal{I}.
\label{eqn:QOptFunc}
\end{equation}
\textbf{To prove:} The truncated Q-function of the mediator is a fixed point of a contraction operator for each objective $Q_{\rho_i}$, ensuring that the multi-objective Q-function will converge to the optimal values based on $\phi$, and thus can be found with Q-learning.\\  
\textbf{Definition 2.} Given the mediator's MDP defined by agents' policies $\pi_\mathcal{I}$, the truncated Bellman optimality operator $\bar{T}_{\mathcal{I}_{i}}$ for objective $i$ is defined by
$$
\left(\overline{\mathcal{T}}_{\mathcal{I}_{i}} \bar{Q}_{\rho_i}\right)(s, a_\rho)=r_i(s, a_\rho)+\gamma \mathbb{E}_{s' \sim \mathcal{P}(s, \boldsymbol{a} \mid \pi_\mathcal{I}) } \max _{a'_\rho \mid \phi}\left[\mathbb{E}_{s'_{r_i} \sim  \mathcal{H} }(\tilde{r}_i((s',s'_{r_i}),a_\rho')+\bar{Q}_{\rho_i}\left(s', a'_\rho\right)\right)],
$$
where $Q_{\rho_i}$ represents the expected returns for an agent $i \in \mathcal{I}$ as estimated by the mediator and is an element of a vector space $\mathcal{Q}_{S A}=\{S \times A \rightarrow \mathbb{R}\}$.\\
\textbf{Lemma 1.} Operator $\overline{\mathcal{T}}_{\mathcal{I}_{i}}$ is a contraction in the sup-norm. \\
\textbf{Proof.} Let $\bar{{Q}}_{\rho_i}^1, \bar{{Q}}_{\rho_i}^2 \in \mathcal{Q}_{S A}, \gamma \in[0,1)$. The operator $\overline{\mathcal{T}}_{\mathcal{I}_{i}}$ is a contraction in the sup-norm if it satisfies $\left\|\overline{\mathcal{T}}_{\mathcal{I}_{i}} \bar{Q}_{\rho_i}^1-\overline{\mathcal{T}}_{\mathcal{I}_{i}} \bar{Q}_{\rho_i}^2\right\|_{\infty} \leq \gamma\left\|\bar{Q}_{\rho_i}^1-\bar{Q}_{\rho_i}^2\right\|_{\infty}$. This inequality holds because:

\[
\begin{aligned}
\left\|\overline{\mathcal{T}}_{\mathcal{I}_{i}} \bar{Q}_{\rho_i}^1-\overline{\mathcal{T}}_{\mathcal{I}_{i}} \bar{Q}_{\rho_i}^2\right\|_{\infty}= & \max _{s, a_\rho\mid \phi} \Big|  \gamma \mathbb{E}_{s'}\left[\max _{a'_\rho\mid \phi}\left(\mathbb{E}_{s'_{r_i}} {\tilde{r}'_i} +\bar{Q}_{\rho_i}^1\left(s', a'_\rho\right)\right)-\max _{a'_\rho\mid \phi}\left(\mathbb{E}_{s'_{r_i}} {{\tilde{r}'_i}} +\bar{Q}_{\rho_i}^2\left(s', a'_\rho\right)\right)\right]+{r}_i(s, a_\rho)-{r}_i(s, a_\rho) \Big| \\ & \leq  \max _{s, a_\rho\mid \phi}\left|\gamma \mathbb{E}_{ s'} \max _{a'_\rho\mid \phi} \mathbb{E}_{s'_{r_i}} \left[ {\tilde{r}'_i} +\bar{Q}_{\rho_i}^1\left(s', a'_\rho\right)- {\tilde{r}'_i} -\bar{Q}_{\rho_i}^2\left(s', a'_\rho\right)\right]\right| \\ & = 
 \max _{s, a_\rho\mid \phi}\left|\gamma\mathbb{E}_{s'} \max _{a'_\rho\mid \phi}\left[\bar{Q}_{\rho_i}^1\left(s', a'_\rho\right)-\bar{Q}_{\rho_i}^2\left(s', a'_\rho\right)\right]\right| \\ & \leq  \max _{s, a_\rho\mid \phi} \gamma \max _{s', a'_\rho\mid \phi}\left|\bar{Q}_{\rho_i}^1\left(s', a'_\rho\right)-\bar{Q}_{\rho_i}^2\left(s', a'_\rho\right)\right|=\gamma\left\|\bar{Q}_{\rho_i}^1-\bar{Q}_{\rho_i}^2\right\|_{\infty}
\\ 
\end{aligned}
\] 
Because $\overline{\mathcal{T}}_{\mathcal{I}_{i}}$ is a contraction as shown above, by the Banach theorem, it admits a unique fixed point $\bar{Q}_{\rho_i}^*$ s.t. $\forall s, a_\rho: \bar{Q}_{\rho_i}^*(s, a_\rho)=\left(\overline{\mathcal{T}}_{\mathcal{I}_{i}} \bar{Q}_{\rho_i}^*\right)(s, a_\rho)$. We now show that this fixed point is the truncated Q-function. Define $Q_{\rho_i} ((s,s_{r_i}), a_\rho)= \mathbb{E}_{s_{r_i}}  s_{r_i}  +\bar{Q}_{\rho_i}^*(s, a_\rho)$. Notice that the fixed point satisfies:
\[
\begin{aligned}
\forall s \in S, a_\rho \in A_\rho: & \bar{Q}_{\rho_i}^*(s, a_\rho)=\left(\overline{\mathcal{T}}_{\mathcal{I}_{i}} \bar{Q}_{\rho_i}^*\right)(s, a_\rho) \\
& =  r_{i}(s, a_\rho)+\gamma\mathbb{E}_{ s'} \max _{a'_\rho\mid \phi}\left[\mathbb{E}_{s'_{r_i}} \tilde{r}'_{i}+\bar{Q}_{\rho_i}^*\left(s', a'_\rho\right)\right] \\
& =  r_{i}(s, a_\rho)+\gamma \mathbb{E}_{s'} \max _{a'_\rho\mid \phi} Q_{\rho_i} \left((s',s'_{r_i}), a'_\rho\right)
\end{aligned}
\]
At the same time, by definition:
\[
\forall s \in S, a_\rho \in A_\rho: \bar{Q}_{\rho_i}^*(s, a_\rho)=Q_{\rho_i}((s, s_{r_i}),a_\rho)- s_{r_i}
\]
Combining the above two equations and swapping terms:
\[
\forall s \in S, a_\rho \in A_\rho: \quad Q_{\rho_i}((s,s_{r_i}), a_\rho)=r_{i}(s, a_\rho)+\mathbb{E}_{ s', s'_{r_i}}\left[s_{r_i}+\gamma \max _{a'_\rho\mid \phi} Q_{\rho_i} \left((s', s'_{r_i}) , a'_\rho\right)\right]
\]
Notice that the last equation shows that $Q_{\rho_i}$ is the fixed point of the Bellman optimality operator $T_{\mathcal{I}_i}$ \eqref{eqn:BellmanOptOp}. i.e., $Q_{\rho_i}=Q_{\rho_i}^*(\pi_\mathcal{I})$, as it satisfies the optimality equation \eqref{eqn:QOptFunc}. It follows that $Q_{\rho_i}^*((s,s_{r_i}), a_\rho \mid$ $\pi_\mathcal{I})= s_{r_i} +\bar{Q}_{\rho_i}^*(s, a_\rho)$, and thus $\bar{Q}_{\rho_i}^*$ satisfies the definition of the truncated Q-function \eqref{eqn:truncatedQ}. The truncated Q-function is then a fixed point of a contraction operator and can be found with Q-learning. 


\section{Additional Experimental Details}
\subsection{Sequential vs. simultaneous learning.}
\label{section:sim_learning}
In this section, we develop simultaneous learning versions of our model and baselines and compare the performance with the sequential version of our model for iterated matrix games. As we see in Figures \ref{fig:simlearn1} and \ref{fig:simlearn2}, both forms of our model, JAM-QL and JAM-QL(sim), converge to similar solutions. 
\begin{figure}[htbp]
  \centering
  \begin{subfigure}{\linewidth}
    \centering
    \includegraphics[width=0.8\textwidth]{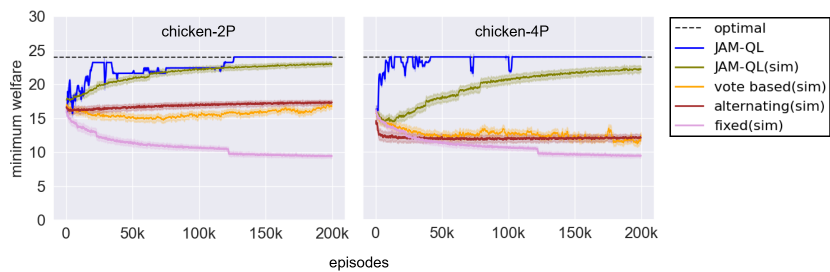}
    \caption{Comparison of minimum welfare through the training runs of JAM-QL with its simultaneous learning version and the simultaneous learning versions of baselines. Results are shown for the \textit{Chicken} game with two players (left) and four players (right). All plots with (sim) represent the simultaneous versions of the models.}
    \label{fig:simlearn1}
  \end{subfigure}

   \vspace{1em} 

  \begin{subfigure}{\linewidth}
    \centering
    \includegraphics[width=0.8\textwidth]{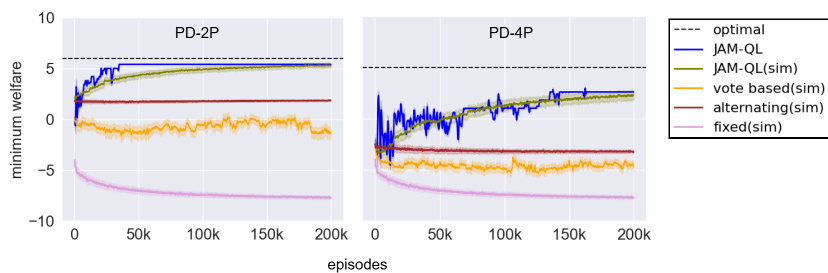}
    \caption{Comparison of minimum welfare through the training runs of JAM-QL with its simultaneous learning version and the simultaneous learning versions of baselines. Results are shown for the \textit{Prisoner's Dilemma (PD)} game with two players (left) and four players (right). All plots with (sim) represent the simultaneous versions of the models.}
    \label{fig:simlearn2}
  \end{subfigure}
\vspace{15pt}
  \caption{Comparison of sequential vs. simultaneous learning.}
  \label{fig:main}
\end{figure}
\vspace{15pt}

\subsection{End-game effects in episodic settings.}
\label{section:end_game_effects}

\begin{figure}[htbp]
  \centering
  \begin{subfigure}{\linewidth}
    \centering
    \includegraphics[width=1.0\textwidth]{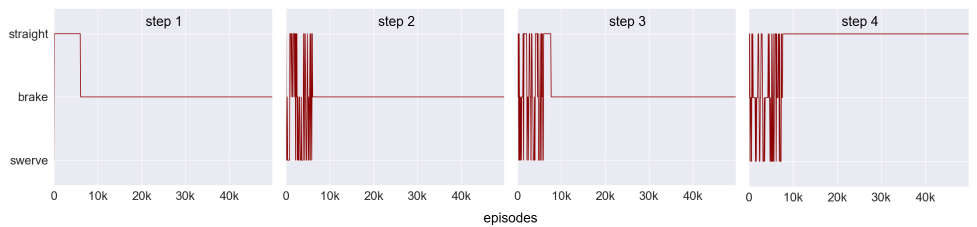}
    \caption{Actions of agent \textit{A}'s leader policy through its training run (50k episodes) for the \textit{Chicken} game without the mediator's additional end-of-game incentive. }
    \label{fig:endgame1}
  \end{subfigure}

  \vspace{1em} 

  \begin{subfigure}{\linewidth}
    \centering
    \includegraphics[width=1.0\textwidth]{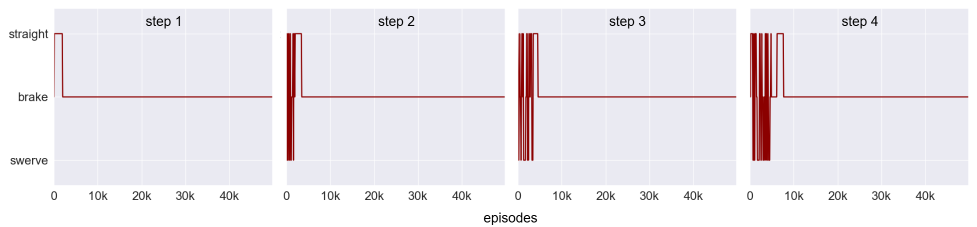}
    \caption{Actions of agent \textit{A}'s leader policy through its training run (50k episodes) for the \textit{Chicken} game with the mediator's additional end-of-game incentive. }
    \label{fig:endgame2}
  \end{subfigure}
\vspace{15pt}
  \caption{End-game effects and the benefits of the mediator's end-of-game stage.}
  \label{fig:main}
\vspace{20pt}
\end{figure}

Here, we analyze the end-game effects that can occur without the mediator's additional end-game incentive for the two-player version of the \textit{Chicken} game defined in the main section. We do this through an experiment where we integrate a simple mediator among two agents, $A$ and $B$, which uses only the historical reward information to determine the leader at any step. Specifically, the mediator selects agent $A$ as the leader if its historical rewards are less than or equal to $B$. Otherwise, it selects agent $B$. Note that the initial leader is selected randomly. We then compare the actions of agent $A$'s leader policy during its course of learning in the presence of a mediator that uses the additional \textit{end-of-game-stage} (stage (iii)) defined in the main text against a mediator that does not. As before, we run each episode of the game for four steps. We show the results in Figures \ref{fig:endgame1} and \ref{fig:endgame2}. 

From Figure \ref{fig:endgame1}, we see that, without the additional end-game incentive, agent $A$ converges to taking the fair action (\textit{brake}) in all steps except the last step (step = 4). This behavior is straightforward to understand. Since it is a leader-advantage game, agent $A$ gains more rewards at any step if it is the leader. To be selected as the leader at any step of the episode, agent $A$ needs to take actions such that its historical rewards in the steps before it are not greater than agent $B$'s, which it can achieve through the fair \textit{brake} action. Hence, it is incentivized to take fair actions in all steps except the last. However, in the final step, it has no further incentive, causing it to take the unfair action (\textit{straight}). In contrast, if the mediator uses an additional end-of-game stage to threaten the agents with a zero-sum transfer of rewards, then agent $A$ has an additional incentive to take the fair action, even at the last step, to prevent the transfer. This can be seen in figure \ref{fig:endgame2}, where the agent $A$ converges to taking the fair action \textit{brake} in all steps of the episode, including the last step, when the mediator uses the additional end-of-game stage.

\subsection{Resource collection with four players.}
\label{section:RC_4P}
We define the four-player version of the resource collection environment here. There are four agents, $A$, $B$, $C$  and $D$, and three types of resources - \textit{red, blue}, and \textit{green}. Agents $A$ \& $C$ ($B$ \& $D$) have the preference and skill to lead and collect \textit{red} (\textit{blue}) resources, while all agents possess the skill to lead and collect \textit{green} resources. However, to collect any resource, a leader needs the help of at least one follower. Collecting \textit{red} or \textit{blue} resources results in unfair returns, favoring the agents who prefer it. Collecting \textit{green} resources yields fair and equal returns. 

We again consider two different settings as in the main text:
\begin{enumerate}
    \item RC-1: Only two types of resources are present - \textit{red}/\textit{blue} (randomized) and \textit{green}, and a maximum of two resources can be collected. As long as leaders with a preference for unfair resources are not selected, fairness can be optimized. 
    \item RC-2: All types of resources are present - \textit{red}, \textit{blue}, and \textit{green}, and a maximum of two resources can be collected. In this case, all agents prefer collecting unfair resources; therefore, fairness
    can only be optimal if additional incentives exist to collect fair resources.  
\end{enumerate}

Collecting an unfair resource yields a reward of \{5, 1\} in favor of all the agents that prefer it, whereas collecting a fair resource yields a reward of \{4, 4\}. All other environmental conditions remain the same as for the two-player version in the main text.\\

\subsection{Implementation details of "vote-based" baseline.}
For the "vote-based" baseline, we augment each agent with an additional learnable policy
purely for voting purposes. Through this policy, each agent chooses one of
the $N$ agents (including itself) as the leader at every leader-selection stage. We
again use Q-learning to learn these voting policies, with the same environmental
reward information used for the action-selection policies.

\section{Hyperparameters}
\label{section:hyperparameters}
We utilize function approximation with deep Q-networks (DQN) for learning the agents' leader policies and the mediator policy for the resource collection environments. We add another DQN network to learn the history model of the mediator.
The neural networks consist of 2
hidden fully connected layers and an output layer. The fully connected layers consist of 128 neurons. ReLU activations follow all hidden layers. The discount factor is set to $\gamma=0.99$ for the mediator and $\gamma=0.9$ for the agents. All networks use Adam optimizer with a learning rate of 0.0001. The following hyperparameters are common to all networks. The exploration rate $\epsilon$ is
initialized at 0.5 and is exponentially annealed to 0.01 throughout the training. The size of the experience replay buffer is 100000
tuples. The batch size is set to
128 transitions, sampled from the replay buffer. The alternating frequency for all frameworks with sequential learning is 100 episodes.

\end{document}